\def\etal{\it et al.\ \rm}
\def\simlt{\hbox{ \rlap{\raise 0.425ex\hbox{$<$}}\lower 0.65ex\hbox{$\sim$} }}
\def\simgt{\hbox{ \rlap{\raise 0.425ex\hbox{$>$}}\lower 0.65ex\hbox{$\sim$} }}

\def\msun { \rm {M_\odot}}

\documentstyle[aas2pp4,epsf,astrobib]{article}
\begin{document}

\title{ First detection of a gravitational microlensing candidate towards 
the Small Magellanic Cloud}

\author{
	  C.~Alcock\altaffilmark{1,2}, 
	R.A.~Allsman\altaffilmark{3},
	  D.~Alves\altaffilmark{1,4},
	T.S.~Axelrod\altaffilmark{3,1},
	  A.C.~Becker\altaffilmark{5,2},
	D.P.~Bennett\altaffilmark{6,1},
	K.H.~Cook\altaffilmark{1,2},
	K.C.~Freeman\altaffilmark{3},
	  K.~Griest\altaffilmark{7,2},
	  M.~J.~Keane\altaffilmark{8},
	M.J.~Lehner\altaffilmark{7,2},
	S.L.~Marshall\altaffilmark{1},
	D.~Minniti\altaffilmark{1},
	B.A.~Peterson\altaffilmark{3},
	M.R.~Pratt\altaffilmark{5},
	P.J.~Quinn\altaffilmark{9},
	A.W.~Rodgers\altaffilmark{3},
	C.W.~Stubbs\altaffilmark{5,2},
	  W.~Sutherland\altaffilmark{10},
	  A.~B.~Tomaney\altaffilmark{5},
	  T.~Vandehei\altaffilmark{7,2},
	  D.~Welch\altaffilmark{11}
	}
\begin{center}
{\bf (The MACHO Collaboration) }\\
\today. 
\end{center}

\vspace{-10mm}
\begin{abstract} 
We report the first discovery of a gravitational microlensing candidate
towards a new population of source stars, the Small Magellanic Cloud (SMC).
The candidate event's light curve
shows no variation for 3 years before an upward excursion lasting
$ \sim 217$ days that peaks around January 11, 1997 at a magnification of
$ \sim 2.1$.
Microlensing events towards the Large Magellanic Cloud and the Galactic bulge
have allowed important conclusions to be reached on the stellar
and dark matter content of the Milky Way.  
The SMC gives a new line-of-sight through the Milky Way, and is
expected to prove useful in determining the flattening of the Galactic halo.

\end{abstract}


\altaffiltext{1}{Lawrence Livermore National Laboratory, Livermore, CA 94550}

\altaffiltext{2}{Center for Particle Astrophysics,
	University of California, Berkeley, CA 94720}

\altaffiltext{3}{Mt.~Stromlo and Siding Spring Observatories,
	Australian National University, Weston, ACT 2611, Australia}

\altaffiltext{4}{Department of Physics, University of California,
	Davis, CA 95616}

\altaffiltext{5}{Departments of Astronomy and Physics,
	University of Washington, Seattle, WA 98195}

\altaffiltext{6}{Department of Physics, University of Notre Dame, 
	Notre Dame IN 46556}

\altaffiltext{7}{Department of Physics, University of California,
	San Diego, CA 92093}

\altaffiltext{8}{ Cerro Tololo Interamerican Observatory, National Optical 
	Astronomy Observatories}

\altaffiltext{9}{ European Southern Observatory, Karl Schwarzschild Str. 2, 
	D-85748 Garching bei Mu\"nchen Germany}

\altaffiltext{10}{Department of Physics, University of Oxford,
	Oxford OX1 3RH, U.K.}

\altaffiltext{11}{Department of Physics and Astronomy, McMaster University,
	Hamilton, ON L8S 4M1, Canada}

\section{Introduction}

Gravitational microlensing has become a new tool for discovering and
characterizing populations of dark objects.  By nightly monitoring of
millions of stars, several groups have detected
the rare brightenings that occur when a dark object passes between a source
star and the observer
\cite{nat93,eros93,ogle93,duo}.  
These events have led to powerful statements on the dark matter content
of the Milky Way 
\cite{lmc1,lmc2,eros},	
strong limits on the possibility of planetary
mass dark matter 
\cite{spike,erosplanets},	
as well as important discoveries
regarding the distribution of mass towards the Galactic bulge
\cite{bulge45,ogletau}.	
See \citeN{pac96}, or \citeN{roulet} for a general review,
and \citeN{beaulieu}, and \citeN{ansari} for discussion of the interpretation
of EROS events.
All previous microlensing results were found using source stars either towards
the Large Magellanic Cloud (LMC) or the Galactic bulge, although
microlensing searches towards M31 have also been
undertaken, and possible events reported 
\cite{crotts}. 
In this paper we announce the first discovery of a microlensing candidate
toward a new target galaxy, the Small Magellanic Cloud (SMC).
The candidate event's light curve
shows no variation for 3 years before an upward excursion lasting
$ \sim 217$ days that peaks around January 11, 1997 at a magnification of
$ \sim 2.1$.

The SMC gives a new line-of-sight
through the Milky Way halo, and a new population of source stars.
Microlensing towards the SMC is important since a comparison of the microlensing
rate towards the SMC with the rate towards the LMC has been predicted to be
a powerful way of measuring the flattening of the Milky 
Way dark halo
\cite{sackett,frieman,explore}. 
Most workers assume that the dark halo is spherical, but there is little
observational or theoretical reason to believe this is so.
A direct 
measurement therefore would be very valuable.  In addition, new lines-of-sight
allow for discrimination between various theories
\cite{zhao2}	
for the populations responsible for LMC microlensing.  A dwarf galaxy 
or possible stellar stream between us and
the LMC is unlikely to also be responsible for 
microlensing towards the SMC.  
In addition, since initial microlensing results towards both the LMC and bulge
were surprising, the potential for initial surprises from SMC microlensing 
is large.

\section{Data}

The MACHO microlensing survey employs a dedicated telescope on Mount
Stromlo, Australia with two simultaneous red and blue passbands and a 
0.5 square degree field of view.  
See 
\citeN{lmc1} 
and references therein
for details of the telescope, camera, photometry, data base, 
and analysis systems.
We have published
results on 22 LMC fields
\cite{lmc1,lmc2,spike}	
containing on the order of 
8.5 million stars over a period of 2 years and
our  four-year analysis of 30 LMC fields is currently underway.
For the SMC we have monitored 21 fields, 6 with good temporal
coverage, for a period of 4 years.  These top 6 fields in the SMC contain
approximately 2.2 million stars.  
Exposures of 300 seconds were taken for all LMC and early SMC images,
but due to crowding and distance they 
were subsequently increased to 600 s for the SMC.  
Approximately 2/3 of the over 4400 observations of the top 6 SMC fields
are 600 s.  Altogether, SMC observations span the dates May 15, 1993 
to the present,
giving an average of 730 observations per field with a mean sampling 
of about 2-3 days.

For the SMC, photometry over our entire four-year data set has not yet been
completed, but after running our alert software on over two years of
photometry, possible microlensing candidates, as well as possible
supernova and nova candidates were found.  Figure 1 shows
the photometric lightcurve for the best microlensing candidate found by 
this method.  The source star 
\renewcommand{\thefootnote}{\fnsymbol{footnote}}
in this event is located 
at $\alpha$ = 01 h 00 min 05.7 s, $\delta$ = -72$^{\circ}$ 15' 01''
(J2000).
A finding chart is available 
at {\tt http://darkstar.astro.washington.edu}.
The star has median magnitudes V = 17.70, and R = 17.66. 
A color magnitude diagram for the area near this star is shown in Figure~2,
indicating that this is an upper main sequence star.

Also shown in Figure 1 is a fit to the theoretical microlensing
light-curve 
(see \citeN{pac86}). 
There are five parameters to the fit: (1,2) the baseline fluxes, (3) the
maximum magnification 
$A_{max}=2.074\pm0.005$, (4) the duration 
(Einstein ring diameter crossing time) $\hat{t} = 216.5\pm1.1$ days, and
(5) the time of peak magnification $ t_0=1836.0\pm0.3$ d or 
roughly January 11, 1997.  
The $\chi^2_{ml}$ to the fit is 2391 with 857 degrees of freedom.
A measure of the signal to noise
\cite{lmc1}	
is given by $\Delta\chi^2/\chi^2_{ml} 
\equiv (\chi^2_{const} - \chi^2_{ml})/\chi^2_{ml}
\approx 30000$.
However, examination of the best seeing images shows that
the source star is 
blended with another SMC star. [An image showing the blend is available at
http://darkstar.astro.washington.edu.]
We performed another fit taking into
account this possibility, in this case finding a best fit blend with 
$\sim 72$\% of the flux in the lensed primary, 
giving $A_{max}=2.5$, $\hat{t} = 247$ days, and an almost identical $t_0$ and
$\chi^2$/dof = 2.77.  Using a stack of good seeing images and an image
obtained from our alert system
\cite{alert},	
we performed independent photometry 
and find the source likely to be a blended star with 77\% of the flux in
the lensed primary.  
Using image differencing techniques
\cite{tomaney}	
to subtract the primary
also gives similar results.
A blended fit with the
blend fraction set to 77\% gives 
$A_{max}=2.4$, $\hat{t} = 242$ days, and again 
$\chi^2$/dof = 2.77.  
The source could contain other undetected blended stars, 
so the above blend fractions
are upper limits, implying a duration of $\simgt 217 - 247$ days.
We note that a centroid shift of a star during lensing
can also give information about blending \cite{duo}, 
but we did not perform this analysis in this case since we
can resolve the stars in some images.

Given the flat baseline, high signal to noise,
the good shape, and the achromaticity we classify this as an excellent
microlensing candidate.  
However, even taking into account the blend, the source star is more 
luminous than
previous LMC events, and near the region of the color-magnitude diagram
susceptible to contamination from ``bumper" type variable stars
\cite{lmc1}.	
Apart from this,
it passes the selection criteria used for LMC microlensing in 
\citeN{lmc2},	
and it also passes selection criteria used for bulge events.

The scatter in the lightcurve around the fit
is larger than the photometry error bars imply, as the fit $\chi^2/$dof
of 2.8 indicates.  This $\chi^2/$dof is somewhat larger than for most of
our LMC microlensing events
\cite{lmc2},	
however, 
the SMC is a new target and we have not yet characterized our photometry
as we have for the LMC and bulge.  The blending also causes systematic
errors in the photometry.  Currently a more refined analysis of 
the data are underway, including a difference frame analysis which may greatly
reduce the problems associated with blending
\cite{tomaney}.	

\section{Discussion}

A comparison of the SMC and LMC microlensing optical depth is very useful.
Almost independent of the dark matter halo model, the ratio
$\tau_{LMC}/\tau_{SMC}$ gives a good measurement of the flattening of
the halo
\cite{sackett,frieman,explore}.	
The optical depth is the probability that a given
star is lensed, and
can be estimated from $\tau_{est} = (\pi/4 E) 
\Sigma {\hat{t}_i}/\varepsilon(\hat{t}_i)$,
where $E$ is the total exposure in star-years, and 
$\varepsilon(\hat{t}_i)$ is the detection efficiency for event $i$
\cite{lmc1}.	

Since we have not yet run our standard analysis to select microlensing
events or the Monte Carlo
needed to calculate $\varepsilon(\hat{t}_i)$, we cannot yet give 
a good estimate of $\tau_{SMC}$.  
For a very rough estimate for this one event alone, 
we can use $E \sim 2.2 \times 10^6$ stars times the 
roughly 850 days in this preliminary search, 
and use $\epsilon \sim 0.3$ to $0.5$ from the LMC
\cite{lmc2}.  
Then with durations quoted above,
we find $\tau \sim 1.5 \times 10^{-7}$ to $3 \times 10^{-7}$, 
consistent with that of our reported
\cite{lmc2} 
LMC optical depth.  Note that this estimate will change when
the search over all the SMC data is completed, when
proper values of $E$ and $\epsilon$ are calculated, and when proper event
selection is performed.

We note that the long duration of this event
($217$ days) would imply a mass of approximately $\sim2.5 \msun$ for
lenses in a standard dark halo with masses $1.0 \msun$ and $10.0 \msun$
being roughly half as likely.  
For a blended fit $\hat{t}=247$ days, the corresponding mass would
be $3.2 \msun$.
However, it is also possible
that both the lens and source belong to the SMC.  In this case,
a wide range of lens masses would be consistent with 
$\hat{t}= 217 - 247$ days and internal velocities 
$v \approx 30$ km/s for the SMC,
depending upon the line-of-sight depth of the SMC.
The line-of-sight depth of the SMC is relevant to
the interpretation of observed microlensing events, and has been studied
by several authors
\cite{mathewson,welchphd,welchetal,caldwell}.	
We conclude from 
\citeN{caldwell}	
and 
\citeN{welchetal} 
that the extreme line-of-sight depth at any position
in the SMC is unlikely to be greater than 12 kpc and that in
field 207 (in the north-central region of the SMC) is unlikely to
be greater than 6 kpc and possibly much less.
Thus, naively, one does not expect the optical depth from SMC/SMC lensing
to be large enough to account for even one such event.
A more complete analysis will be presented elsewhere after photometry
and analysis of the SMC fields have been completed.

\acknowledgements

We are grateful for the support given our project by the technical
staff at the Mt. Stromlo Observatory.  Work performed at LLNL is
supported by the DOE under contract W-7405-ENG-48.  Work performed by the
Center for Particle Astrophysics personnel is supported by the NSF
through AST 9120005.  The work at MSSSO is supported by the Australian
Department of Industry, Science, and Technology. K.G. and M.J.L. acknowledge
support from DOE, Alfred P. Sloan, and Cottrell Scholar awards.
C.S. acknowledges the generous support of the Packard, Seaver, and 
Sloan Foundations.
W.S. is supported by a PPARC Advanced fellowship.

\newpage
\figcaption[]{
The observed light-curve for 207.16604.214,
with estimated $\pm 1\sigma$ errors. 
The upper panel shows $A_{blue}$, the flux (in linear units) 
divided by the median 
observed baseline flux, in the blue passband. 
The lower panel is the same for the red passband.  
The smooth curve shows the 
best-fit theoretical (non-blended) microlensing model, fitted
simultaneously to both colors.  Some data points are missing from the red
lightcurve due to a dead area on one CCD.
\label{figlc}}

\figcaption[]{
Color-magnitude diagram of stars within roughly 5 arc minutes of 
the SMC microlensing candidate.  
The magnitudes are approximately Kron-Cousins $V$ and $R$.
The candidate is marked as a large dot.
\label{figcmd}}

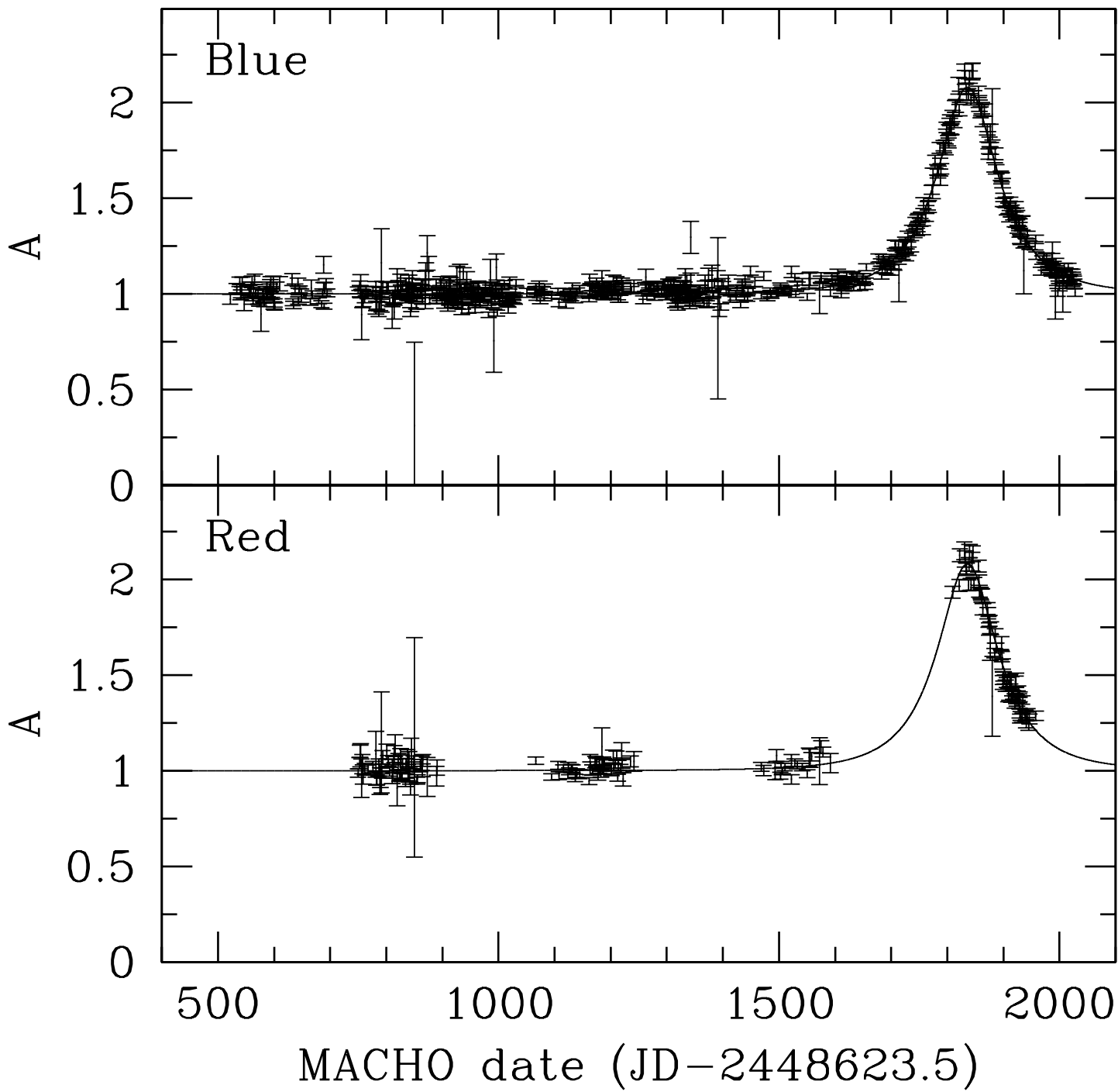
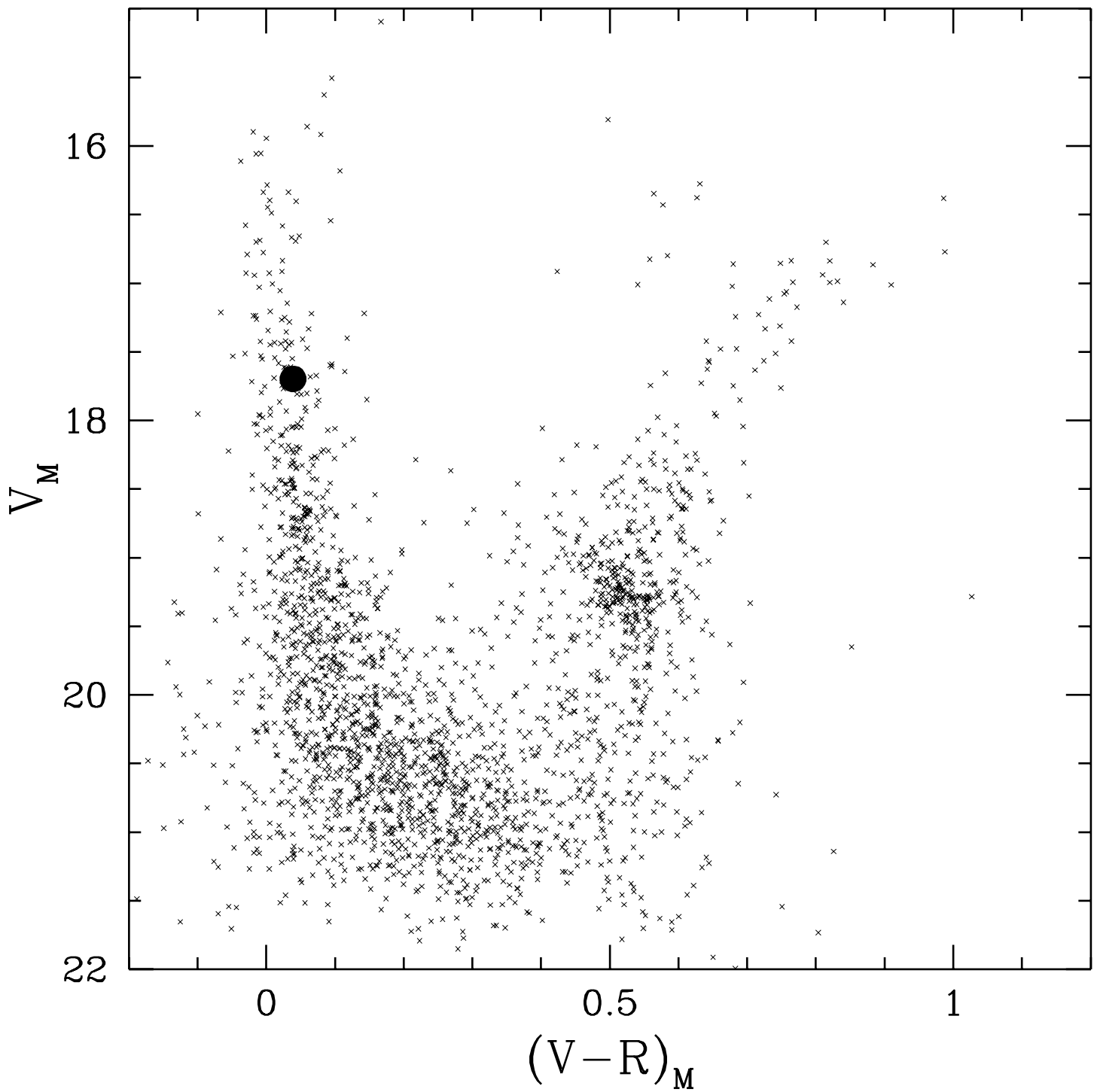
\end{document}